\documentclass[aps,amsmath,amssymb,12pt,showpacs,nofootinbib]{revtex4}
\usepackage[T2A]{fontenc}
\usepackage[utf8]{inputenc}
\usepackage[english,russian]{babel}
\usepackage{graphicx}
\usepackage{hyperref}

\begin{document}

\selectlanguage{english}
\title{The relative yields of heavy hadrons as function of transverse momentum at  LHC experiments}

\author{\firstname{A.~V.}~\surname{Berezhnoy}}
\email{Alexander.Berezhnoy@cern.ch}
\affiliation{SINP MSU, Moscow, Russia}

\author{\firstname{A.~K.}~\surname{Likhoded}}
\email{Anatolii.Likhoded@ihep.ru}
\affiliation{IHEP, Protvino, Russia}

\begin{abstract}
It is shown that the dependence of ratio $f_s/f_d$  could be understood within the fragmentation approach.
However, the validity of the fragmentation model is questionable at low $p_T$, and  contributions of nonfragmentation mechanisms are possible. 
It is demonstrated, that the precise measurements of the $p_T$ dependence of $f_s/f_d$, $f_{\Lambda_b}/f_{B}$ and analogous values as function of  $p_T$  can essentially improve our understanding of heavy hadron production.
At large $p_T$ a plateau is expected  in the $f_s/f_d$  distribution. Contrary, a plateau is not expected in  the $f_{B_c}/f_{B}$  distribution as a function of $p_T$.
\end{abstract}


\maketitle

\section{Introduction}

The commonly applied method in the calculation of heavy hadron production cross section is based on the factorization theorem. 
Within this approach the production process is subdivided into the perturbatively calculated hard production of heavy quark and the soft process of heavy quark hadronization.  The hard part can be calculated within several methods:  FONLL~\cite{Cacciari:1998it,Cacciari:2012ny}, pQCD  LO + $k_T$-factorization~\cite{Baranov:2005xt}, GM-VFNS~\cite{Kniehl:2012ti}. The hadronization this framework is completely determined by the so-called fragmentation function (FF), which can be extracted from the data on   $B$-meson production in  $e^+e^-$-annihilation.

This approach is not valid for the $B_c$-meson hadronic  production at low and medium transverse momenta. As it is shown in our  analysis~\cite{Berezhnoy:1994ba,Berezhnoy:1995au,Berezhnoy:1997fp},  the fragmentation satisfactorily describes  hadronic production only at $B_c$ transverse momenta larger than $\sim 40~\mathrm{GeV}$ (see also~\cite{Kolodziej:1995nv, Chang:1994aw, 
Baranov:1997sg, Baranov:1997wv}).  Contrary to this, in the $e^+e^-$-annihilation the fragmentation approach could  be applied in the total phase space~\footnote{However, there is no $e^+e^-$ experiment, which allows to detect $B_c$ mesons.} This difference is due the fact,  that  in the  $pp$-interaction  $b$-quarks strongly  with the hadronic remnant. The contribution of such interaction to the $B_c$ production cross section depends on  $p_T$ as $\sim 1/{p_T^6}$, whereas the fragmentation contribution varies as $\sim 1/{p_T^4}$. Obviously, at very high $p_T$ the nonfragmentation contribution is negligible. Nevertheless  at the experimentally observed transverse momenta the nonfragmentation mechanism is dominant for the $B_c$ production and, in principle, could essentially  contribute to the yields of other $B$ mesons. 

Unfortunately, the problem of nonfragmentation contribution is poorly studied. This situation is absolutely unacceptable because the understanding of the production process is strongly needed for estimating branching ratios of beauty mesons at LHC experiments.

As an example, it is worth to mention the notable observation of rare decay $B_s \to \mu^+ \mu^-$ at LCHb~\cite{Aaij:2012nna,Aaij:2013aka} and CMS~\cite{Chatrchyan:2013bka}. To estimate the branching ratio of this decay the experimental ratio of $B_s$  meson and $B_d$ meson yields  has been used.

In this paper we show that the relative yields  of beauty hadrons depends on the transverse momentum, and can not be assumed constant, as is commonly supposed.

\section{Fragmentation}

It is well known that the cross section of heavy quark production can be calculated in the framework of pertubative QCD. The leading order  formula for the heavy quark production is known from \cite{Gluck:1977zm,Combridge:1978kx}. However, it seems that LO can not describe the heavy quark production at CDF and LHC and accounting of NLO~\cite{Mangano:1991jk}, NNLO terms or even infinite sets of terms are needed.  The research involved for such calculations face the  following problems:
  \begin{itemize}
\item how to take into account final and initial state gluonic radiation (DGLAP);
\item how to take into account the mass of heavy quarks (DGLAP is applicable for massless partons);  
\item how to avoid double counting.
\end{itemize}
These problems have no precise and unique solutions. This is why there are several approaches to the problem of cross section calculation within the LO QCD. Some of them are itemized below:  
\begin{itemize}
\item FONLL: NLO (massive quark)  + resummation of large logs (
at $p_T<5m_c$ NLO works without  logarithm resummation)~\cite{Cacciari:1998it}.
\item $k_T$ factorization approach:  LO (massive quark) + virtual initial gluons (it seems that sea $c$ quark is not needed)~\cite{Baranov:2005xt}.
\item General-mass variable-
flavor-number scheme (GM-VFNS): The heavy quark is treated as any other massless parton, the mass is taken into account as large logarithms $\ln(p_T/m)$ in parton distribution and
fragmentation functions, where they are resummed by imposing DGLAP evolution~\cite{Kniehl:2012ti}.
\end{itemize}

It is very convenient to think about the heavy hadron production in term of quarks, but due to the confinement phenomenon isolated quarks are not observed. Only  hadrons can be observed experimentally. Therefore we need somehow to transform the quark cross section into the hadronic cross section,  and fragmentation is a simple way to achieve this.  The main idea of this approach is to suppose, that the heavy quark $Q$ with three momentum $\mathbf{p}$ transforms to the heavy hadron $H$ with  three momentum $z\mathbf{p}$ ($0<z<1$) with the process independent probability $D(z) dz$:
$$Q ( \mathbf{p}) \xrightarrow{ D(z) dz} H(z\mathbf{p}).$$

In the $e^+e^-$ annihilation at high energy the final heavy quarks are monochromatic and the fragmentation approach can be applied in the total phase space.  In such case the fragmentation function can be directly extracted from the experimental data according to the formula
\begin{equation}
\frac{d \sigma_H}{d z} = \sigma_{Q}  D(z),
\end{equation}
which easily can be obtained from the more general formula:

\begin{equation}
\frac{d^2\sigma_H}{d p_T d z}  = \int \delta(p_T-z p_T^Q) D(z) \frac{d\sigma_Q ( p_T^Q)}{dp_T^Q}  d p_T^Q
\label{eq:frag_general}
\end{equation}

The cross section distribution on transverse momentum can also be obtained from (\ref{eq:frag_general}):
\begin{equation}
\frac{d \sigma}{d p_T} = \int^1_{\frac{2 p_T}{\sqrt{s}}} \frac{d \sigma_Q }{d k_T}(p_T/z) \frac{D(z)}{z} d{z}.
\label{eq:frag_pt}
\end{equation}

In the  hadronic interaction heavy quarks are produced with different energies and 
therefore the value of $z$ can not be determined experimentally. In spite of that, the formula (2) for the $p_T$ distribution is applicable for the hadronic interactions and can be tested at experiments.

As variants, instead of $z=\frac{\mathbf{p}_H}{\mathbf{p}_Q}$ another definitions can be used for $z$: 
$z= \frac{E_H}{E_Q}$ or $ z=\frac{\mathbf{P}_H+E_H}{\mathbf{P}_Q+E_Q }$.

For the heavy hadron production all these schemes should work well for $\mathbf{p} \to \infty$ giving practically the same predictions.  Difficulties become essential  at low and medium transverse momenta 
($\mathbf{p} \lesssim 5\div 6 M_Q$):
\begin{itemize}
\item $m_Q\neq m_H$, and therefore the invariant mass can not  be conserved;
\item the predictions depend essentially on the choice of fragmentation variable   ($\mathbf{p}$, $E$, $p_+$);
\item the predictions depend on the coordinate system.
\end{itemize}
 
The main source of the itemized above problems is that within the fragmentation approach the two-particle phase space is applied  instead of the multi particle one.   

It should be noted, that discussing properties of FFs in this paper we will talk mainly about the so-called    
non-perturbative part of FF.  The perturbative part of FF is the same for heavy hadrons with the same flavor of constituent heavy quark. This is why we consider the perturbative FF  part as part of the heavy quark cross section.

Unfortunately, we do not understand clearly the hadronization process for heavy mesons and this is why
the parameters of FFs as a rule must be extracted from experimental data on $e^+e^-$-annihilation.
A case of double heavy mesons is the only exception, where both the shape and the normalization can be predicted within pQCD. For $S$-wave states of $B_c$-mesons the following expressions were obtained in~\cite{Clavelli:1982hp,Ji:1986zr,Amiri:1986zv,Chang:1991bp,Chang:1992bb,Braaten:1993mp,Kiselev:1994qp,
Braaten:1994bz} :
\begin{multline}
       D_{\bar b \to B_c^*}(z)=\\
\frac{2\alpha^2 |R_S(0)|^2}{81\pi m_c^3}
\frac{rz(1-z)^2}{(1-(1-r)z)^6}
(6-18(1-2r)z+(21-74r+68r^2)z^2
\\
-2(1-r)(6-19r+18r^2)z^3+
3(1-r)^2(1-2r+2r^2)z^4)
\label{eq:vect_frag}
\end{multline}
\begin{multline}
D_{\bar b \to B_c}(z)=\\
\frac{2\alpha^2 |R_S(0)|^2}{27\pi m_c^3}
\frac{rz(1-z)^2}{(1-(1-r)z)^6}
(2-2(3-2r)z+3(3-2r+4r^2)z^2
\\
-2(1-r)(4-r+2r^2)z^3+
(1-r)^2(3-2r+2r^2)z^4),
\label{eq:ps_frag}
 \end{multline}

where $r=\frac{m_c}{m_c+m_b}$ and $R_S(0)$ is the value of the $B_c$ nonrelativistic wave function at origin.

\begin{figure}
 \centering
\resizebox*{0.8\textwidth}{!}{\includegraphics{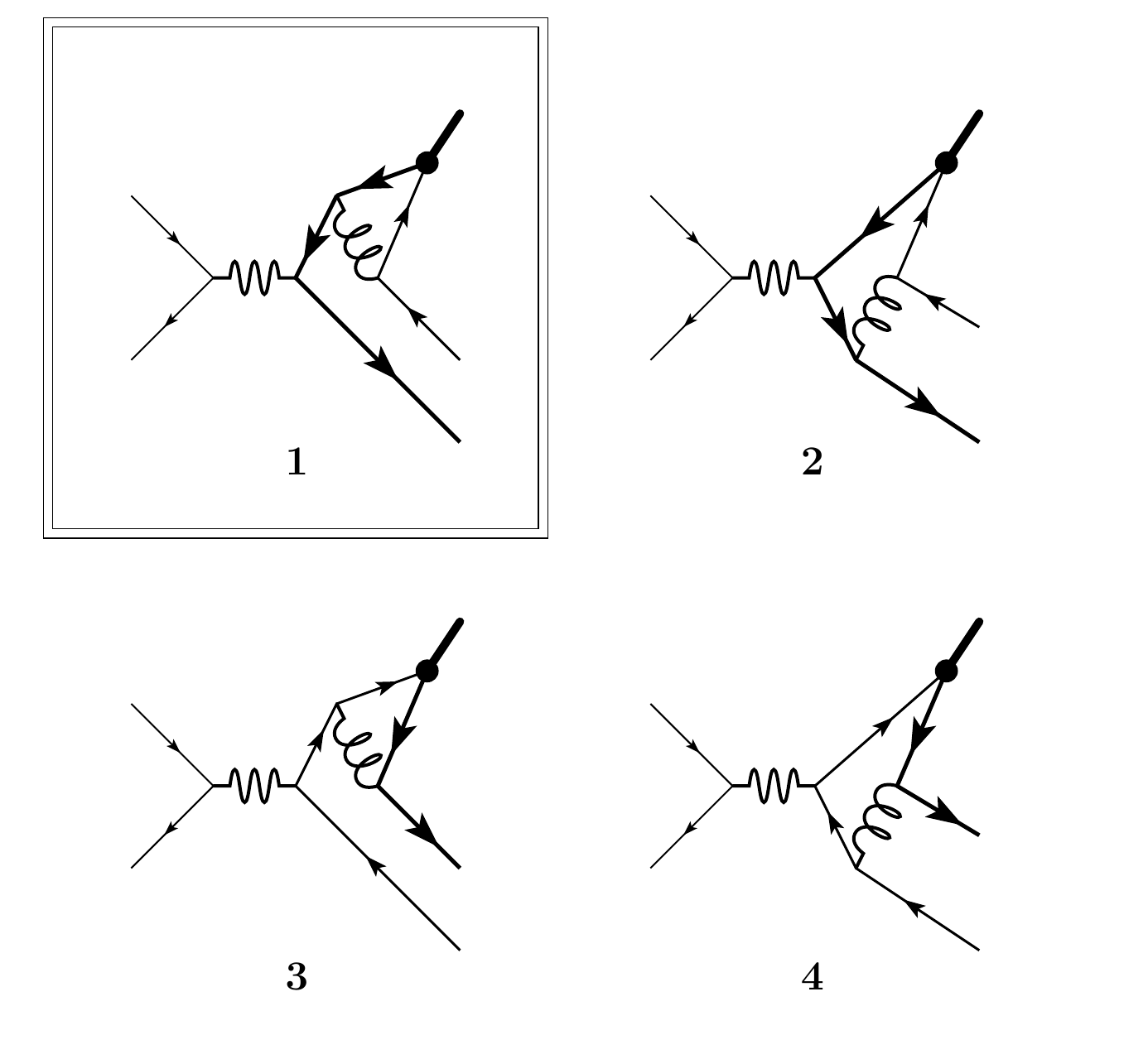}}
\caption{
Feynman diagrams for $e^+e^- \to B_c+X$.}
\label{fig:diagrams_ee_to_BcX}
\end{figure}

The diagrams for this process are shown in Fig.~\ref{fig:diagrams_ee_to_BcX}. The special choice of gauge allows to suppress all diagrams except diagram 1, which can be naturally interpreted as a diagram corresponding to the fragmentation(see~\cite{Braaten:1994bz}).

It is seen from these expressions for $B_c$  $S$-waves states, as well as for $P$-wave~\cite{Yuan:1994hn,Cheung:1995ye} and $D$-wave~\cite{Cheung:1995ir} excitations,  that the shape of the fragmentation function depends on the quantum numbers of final doubly heavy mesons.  We should especially stress here that there are no reasons to assume that for the heavy-light mesons such dependence is negligible.  

The parameterizations  (\ref{eq:vect_frag}) and (\ref{eq:ps_frag})   are also successfully used to describe  the  heavy-light meson production. In this case $r$ is not the ratio between the light quark mass and  
the meson mass, but a phenomenological parameter, which is obtained form a fit to the data. The normalization also can not be theoretically predicted and is one more phenomenological parameter within the discussed model.
 
 The so-called Peterson parametrization~\cite{Peterson:1982ak} has the form
\begin{equation}
D_{Q\to (Q\bar q)} (z)\sim \frac{1}{z\left(1-\frac{1}{z}-\frac{\epsilon}{1-z}
\right )^2}.
\label{eq:peterson_frag}
\end{equation}

The main dependence on $z$ in this parametrization can  be obtained within simple
quantum-mechanical considerations. This dependence is determined by the perturbative propagator of heavy quark:
\begin{equation}
\frac{1}{m_Q^2-p_{Q^*}^2},
\end{equation}
where $p_{Q^*}$ is the momentum of the virtual heavy quark just before transforming to the heavy hadron.
Expanding the denominator of this expression in a rapidly moving
reference frame in terms of the small parameters
$\frac{m_Q}{E_H}$ and $\frac{m_H-m_Q}{m_Q}=\frac{\Delta m}{m_Q}$ and taking into account that $z=\frac{E_H}{E_{Q^*}}$ one obtains:
\begin{equation}
\frac{1}{m_Q^2-p_{Q^*}^2}\sim \frac{1}{1-\frac{1}{z} -\frac{{\Delta m}^2}{m_Q^2}\frac{1}{1-z}} .
\label{eq:propag}
\end{equation}

\begin{figure}[!t]
\centering
\resizebox*{1.0\textwidth}{!}{
\includegraphics{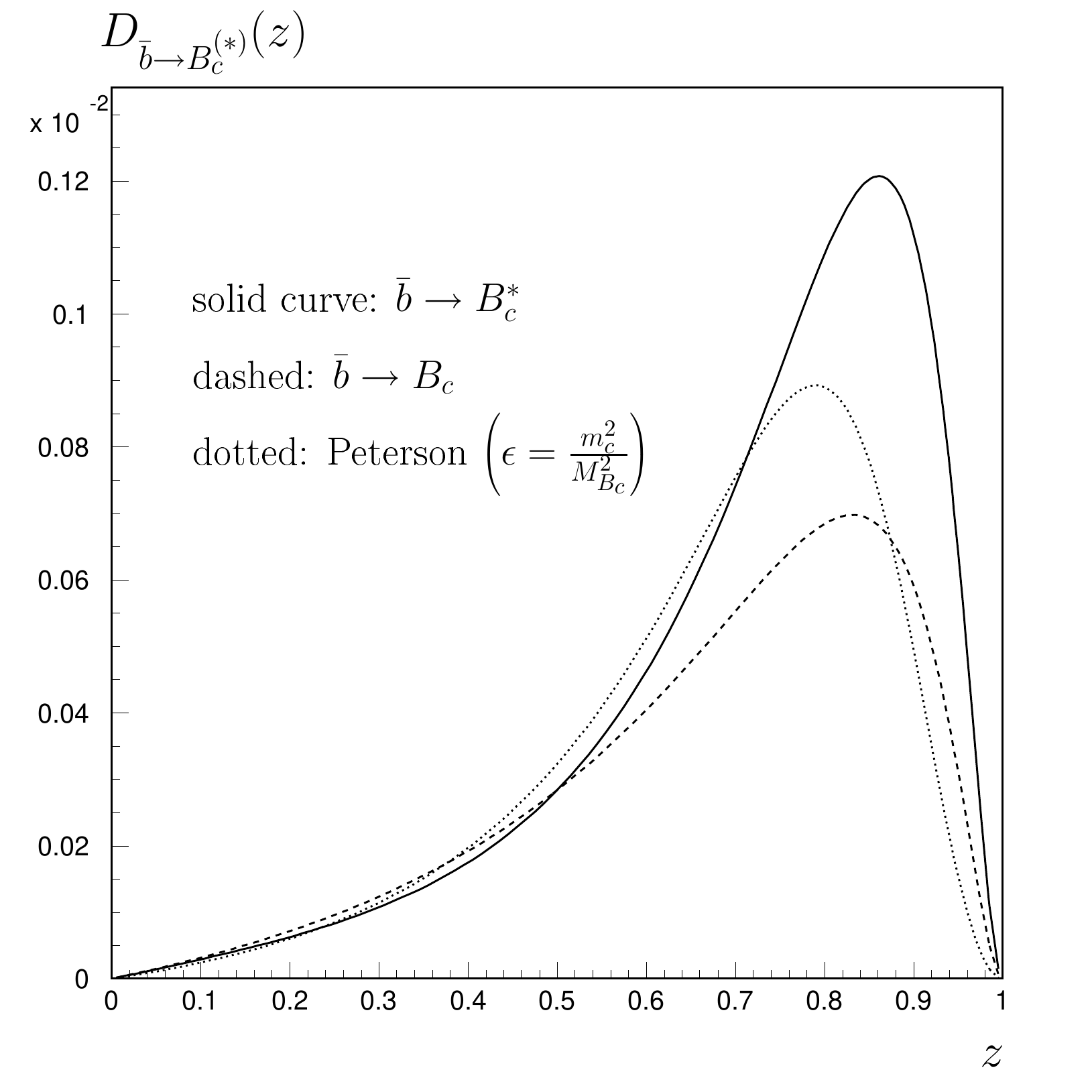}
\hfill
\includegraphics{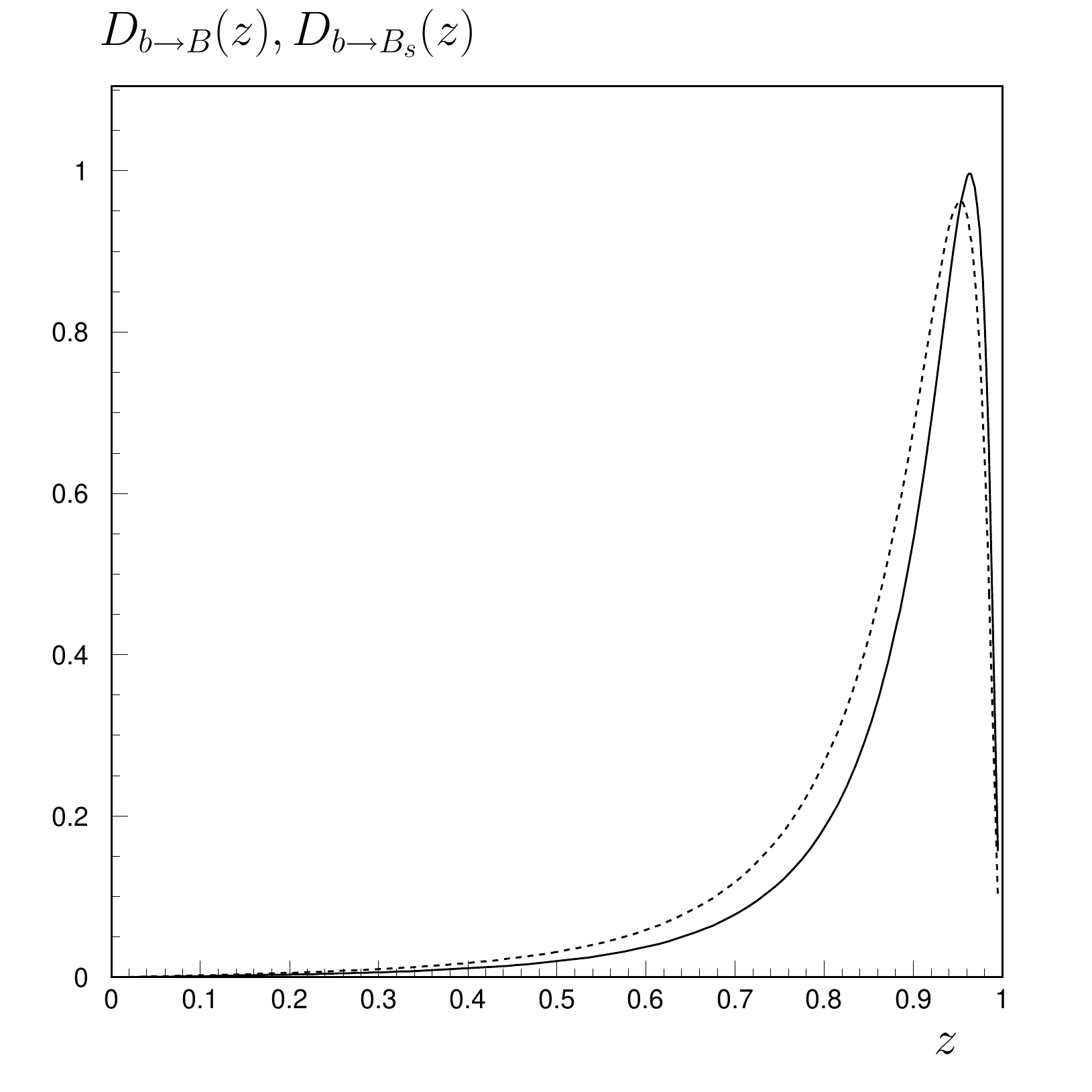}
}
\\
\parbox[t]{0.47\textwidth}{
\caption{FF for $b\to B_c^{(*)}$ obtained within pQCD vs. Peterson parametrization.}
\label{fig:frag_Bc}   \hfill}
\hfill
\parbox[t]{0.47\textwidth}{
\caption{pQCD motivated FF for $B$ (solid) and $B_s$ (dotted).} 
\label{fig:fragbs}   \hfill}
\end{figure}

Thus the main dependence in the Peterson FF can be obtained within perturbative theory, and therefore the Peterson FF can be considered as simplified modification of   pQCD motivated FFs (\ref{eq:vect_frag}) and (\ref{eq:ps_frag}).
Comparing (\ref{eq:peterson_frag}) and (\ref{eq:propag}) one can conclude, that $\epsilon\approx \frac{{\Delta m}^2}{m_Q^2}$. It is interesting to compare pQCD motivated FFs for $B_c$ and $B_c^*$
with Peterson parameterization with $\epsilon=\frac{m_c^2}{M_{B_c}^2}$ (see Fig~\ref{fig:frag_Bc}).

One more parametrization is the Kartvelishvili–
Likhoded–Petrov (KLP) FF~\cite{Kartvelishvili:1977pi}, which is given by a simple power-law
dependence of the form
\begin{equation}
D(z) \sim z^\alpha (1-z)^\beta,
\end{equation}
which was motivated by a similar expression for the
heavy-hadron structure function and by the Gribov–
Lipatov reciprocity~\footnote{The physical meaning of this reciprocity is the following. It is
assumed that the initial-parton wave function contains the
final-hadron wave functions, in just the same way as it is
assumed in considering the structure of hadrons that the
hadron wave function contains the parton wave functions: \\ $| \textrm{\bf hadron}\rangle = \sum_i | \textrm{parton}\rangle_i\quad \Longleftrightarrow \quad 
| \textrm{\bf parton}\rangle = \sum_j | \textrm{hadron}\rangle_j$. }.
Owing to this duality relation, the parameters $\alpha$  and $\beta$ appearing in this expression can be related to the analogous parameters of
the heavy-hadron structure function. The parameter
$\alpha$ is equal to the sign-reversed intercept of the leading Regge trajectory associated with the heavy quark being considered. The other parameter, $\beta$, determines
the asymptotic behavior for $z\to 1$ and, in general,
depends on the sort of spectator quarks in the heavy
hadron and on their number.
Thus, it can be concluded that, in the KLP model,
the behavior of the nonperturbative fragmentation
function for $z \sim 0$ is associated with the heavy quark
 exclusively, while its behavior for $z \sim 1$ is determined by the hadron type.

 By way of example, we indicate that, for heavy quark fragmentation to a non-strange baryon $\beta \sim 3$,
while, for fragmentation to a non-strange meson, $\beta \sim 1$. The parameter $\beta \sim 2.5$ and $\sim 0.5$
for heavy strange baryons and  heavy strange mesons, respectively.
Unfortunately, the model described here does not predict any difference in fragmentation to hadrons having
different angular-momentum values.

In Tab.~\ref{tab:FFs} the different FFs  characteristics are compared to each other. Any approach for heavy quark production can be used  with any 
FF parameterization, but the FF  parameter values essentially depend  on the choice of 
 heavy quark production model. And of course  the cross section of heavy hadrons can only be measured experimentally.

\begin{figure}[!t]
 \centering
\resizebox*{0.6\textwidth}{!}{\includegraphics{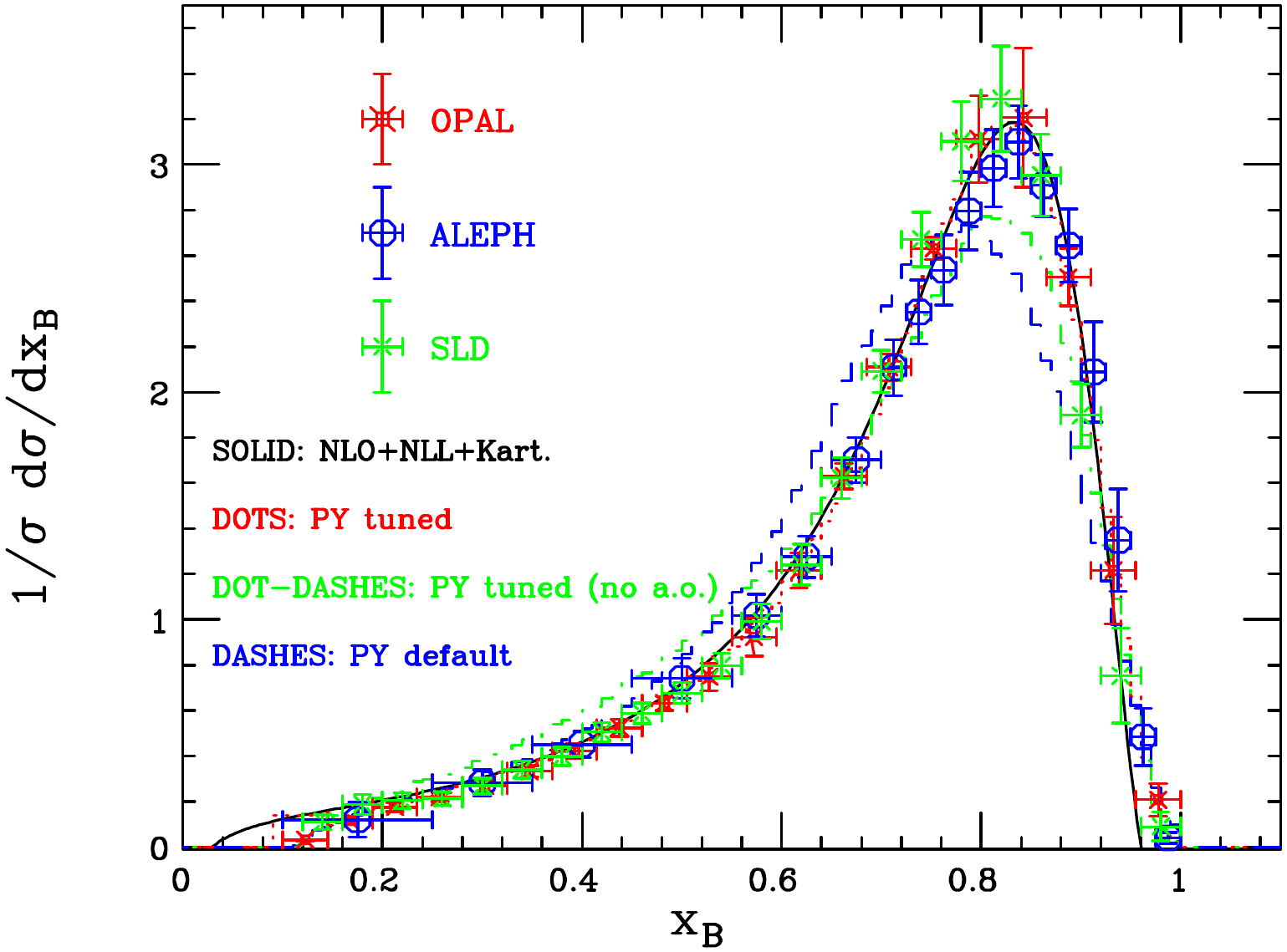}}
\caption{Data from LEP and SLD experiments~\cite{Heister:2001jg,Abbiendi:2002vt,Abe:1999ki}, compared with the NLO+NLL
calculation convoluted with the KLP FF (solid curve).}
\label{fig:KLP_ee_bb}
\end{figure}

For heavy hadron production in $e^+e^-$-annihilation the fragmentation approach works fairly good, as is seen in Fig.~\ref{fig:KLP_ee_bb} taken from~\cite{Corcella:2005dk}, where the data of  the experiments ALEPH, OPAL and SDL~\cite{Heister:2001jg,Abbiendi:2002vt,Abe:1999ki} are compared with the fragmentation approach estimations,  obtained using the KLP approximation.

\begin{table}
\caption{The origin of main dependence on $z$ for different FFs}
\begin{tabular}{|l|l|}
\hline
FF &  the main $z$ dependence\\ \hline
 pQCD motivated & propagator + wave function \\ \hline
  Peterson & propagator  \\ \hline
   KLP & wave function  \\ \hline
\end{tabular}
\label{tab:FFs}
\end{table}

The Peterson function is the most  popular parametrization.  The dependencies on the quantum numbers, as well as on the flavor of the light constituent quark are hidden in this approach, because the features  of this parametrization are determined mainly by the heavy quark propagator.   This might be the reason that it is not always realized, that fragmentation functions for different heavy-light mesons
can have different shapes, even if  all these mesons contains the same type heavy quark. 
Unfortunately, there is a  widespread belief,  that the difference in final meson type leads only to the difference in the relative yields, and the relative yields do not depend on $p_T$. Also there is a belief,  that relative yields at high transverse momenta are equal to the corresponding total relative yields. We will show, that such beliefs are unreasonable.

Indeed, lets us suppose  that two heavy hadrons $H_1$ and $H_2$ are produced in $Q$ quark hadronization with probabilities $f_{H_1}^\mathrm{total}$ and $f_{H_2}^\mathrm{total}$ and show that
\begin{equation}
\frac{d\sigma_{H_1}}{d\sigma_{H_2}}(p_T)= \frac{f_{H_1}}{f_{H_2}}(p_T)\neq \frac{f_{H_1}^\mathrm{total}}{f_{H_2}^\mathrm{total}}
\end{equation}

For the hadronic production of heavy hadrons at high transverse momenta within the fragmentation model we should assume that

\begin{equation}
\frac{d\sigma_H}{dp_T}=\int_{2p_T/\sqrt{s}\approx 0}^1 \frac{d\sigma_{Q}}{d k_T}\left(\frac{p_T}{z}\right)\frac{D(z)}{z}dz \nonumber
\end{equation}
and that at high $p_T$
\begin{equation}
\frac{d\sigma_{Q}}{d k_T}\sim \frac{1}{k_T^4}.
\end{equation}
It is easily to obtain at high  $p_T$ that
\begin{equation}
\frac{d\sigma_H}{dp_T}\sim  \frac{1}{p_T^4}\int_0^1 D(z) z^3dz \sim \frac{d\sigma_{Q}}{d p_T},
\end{equation}
\begin{equation}
\frac{f_{H_1}}{f_{H_2}}(p_T)=\frac{\langle z^3\rangle_{H_1}}{\langle z^3\rangle_{H_2}}\frac{f_{H_1}^\mathrm{total}}{f_{H_2}^\mathrm{total}}=const,
\label{eq:f_ratio}
\end{equation}

where 
\begin{equation}
\langle z^n \rangle=\frac{1}{f_H}\int_0^1  D(z) z^n dz,
\end{equation}

\begin{equation}
\int_0^1  D(z)  dz=f_H.
\end{equation}

 Researchers often omit the factor ${\langle z^3\rangle_{H_1}}/{\langle z^3\rangle_{H_2}}$ in 
 (\ref{eq:f_ratio}) discussing the fragmentation approach. But if  ${\langle z^3\rangle_{H_1}}/{\langle z^3\rangle_{H_2}}\neq 1$ then $\frac{f_{H_1}^\mathrm{total}}{f_{H_2}^\mathrm{total}} \neq \frac{f_{H_1}}{f_{H_2}}(\mathrm{high }\; p_T)$, and therefore ${f_{H_1}}/{f_{H_2}} $ can not be a constant at all transverse momenta.
 
For any power-law $p_T$ dependence of the  quark cross section, the dependence of the hadronic cross section will be the same  and the ratio ${f_{H_1}}/{f_{H_2}}$ will be constant at high transverse momenta: 
\begin{equation}
\frac{d\sigma_{Q}}{d p_T}\sim \frac{1}{p_T^n} \quad \Longrightarrow \quad
\frac{d\sigma_H}{dp_T} \sim \frac{d\sigma_{Q}}{d p_T} \quad \Longrightarrow \quad \frac{f_{H_1}}{f_{H_2}}(\mathrm{high }\; p_T)=const.
\end{equation}

If the $p_T$ dependence is not a "pure" power-law, then   $\frac{f_{H_1}}{f_{H_2}}$ will depend on $p_T$. For example:

\begin{equation}
\frac{d\sigma_{Q}}{d p_T}\sim \frac{1}{p_T^4} +  \frac{a}{p_T^6}\quad \Longrightarrow 
\frac{f_{H_1}}{f_{H_2}}(p_T)\sim 1+
\frac{a}{p_T^2}\left(\frac{\langle z^5\rangle_{H_1}}{\langle z^3\rangle_{H_1}}-\frac{\langle z^5\rangle_{H_2}}{\langle z^3\rangle_{H_2}}\right) 
\end{equation}

If the $p_T$ dependence of the quark cross section has the form ${f_\mathrm{slow}(p_T)}/{p_T^n}$, where $f_\mathrm{slow}(p_T)$ is a slow (for example, logarithmic) function of $p_T$,  then we still can expect
 that ${f_{H_1}}/{f_{H_2}}\approx const$.

It could be concluded from above, that the relative yield of heavy hadrons at  large $p_T$ could depend on $p_T$ due to  power corrections and practically do not depend on logarithmic corrections.

\begin{figure}[!t]
\centering
\resizebox*{1.0\textwidth}{!}{
\includegraphics{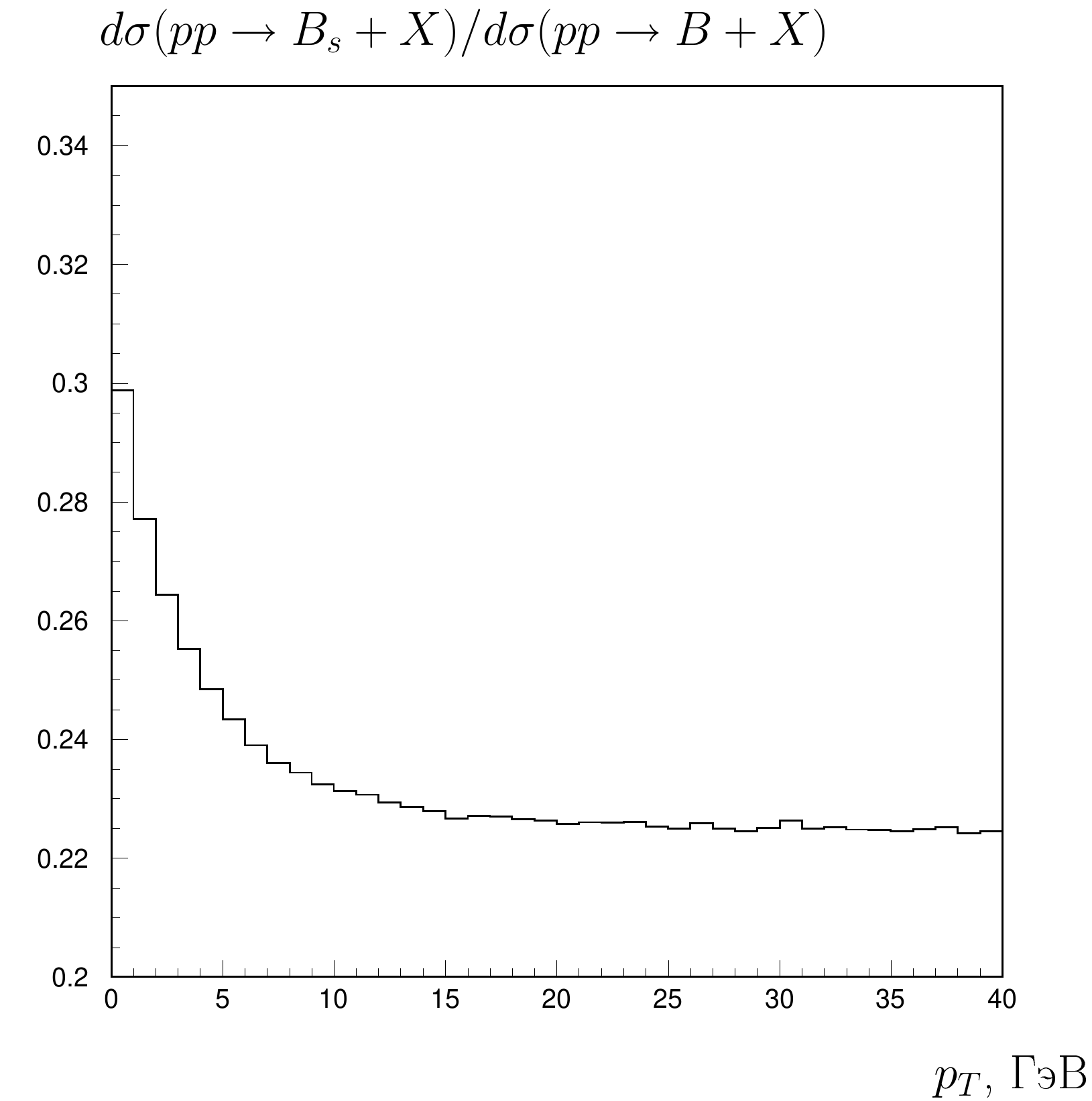}
\hfill
\includegraphics{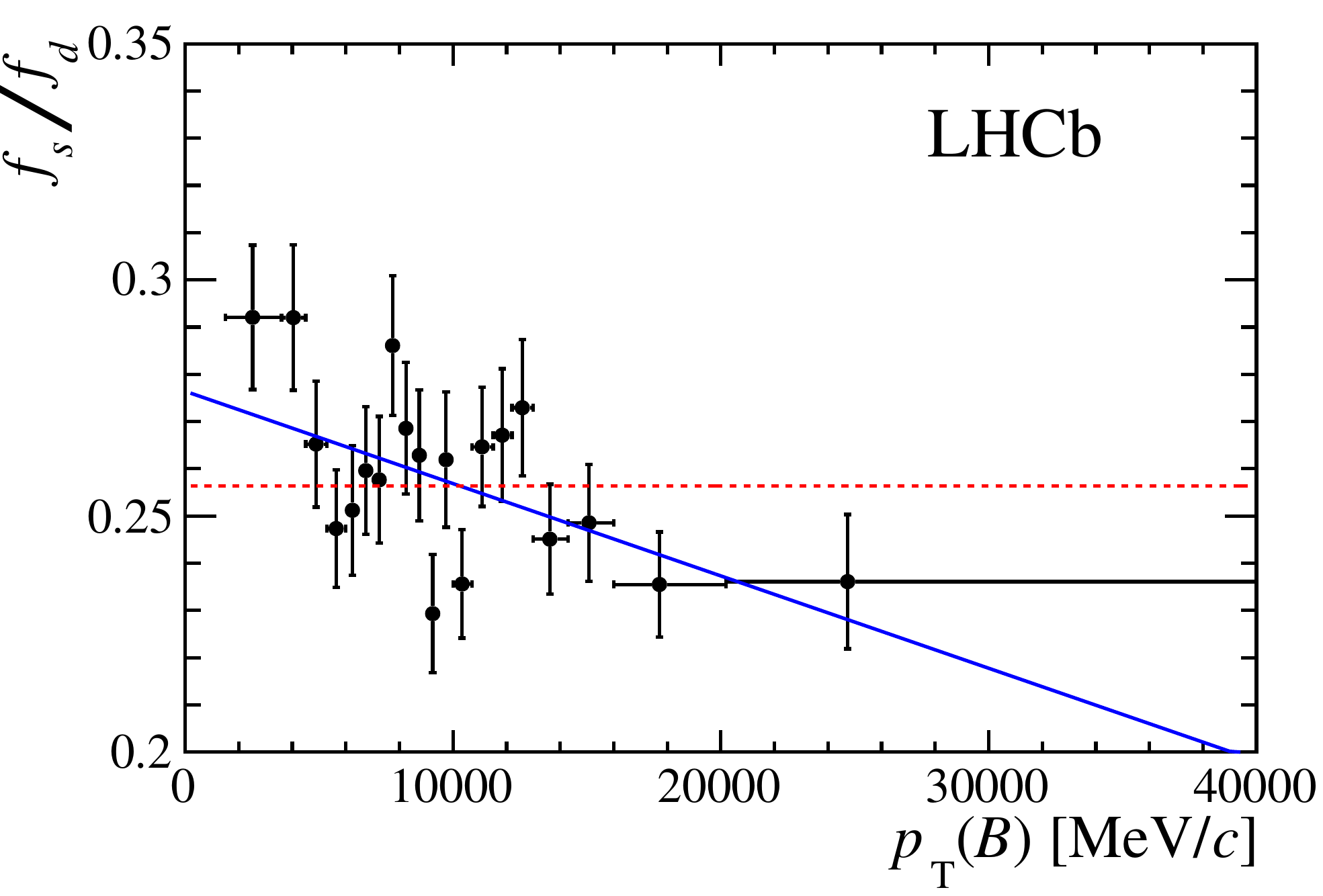}
}
\\
\parbox[t]{0.47\textwidth}{
\caption{$f_s/f_d$ ratio as function of $p_T$}
\label{fig:BsBd}   \hfill}
\hfill
\parbox[t]{0.47\textwidth}{
\caption{$f_s/f_d$ ratio as function of $p_T$ at LHCb.} 
\label{fig:BsBd_LHCb}   \hfill}
\end{figure}

To demonstrate that the dependence of ${f_{B_s}}/{f_{B}}$ on $p_T$ could be explained within the fragmentation model, let us use the FFs for $B$ and $B_s$ mesons in  pQCD motivated parametrizations with $r=0.057$ ($m_b=5.0$~GeV, $m_d=0.3$~GeV) and $r=0.074$ ($m_b=5.0$~GeV, $m_s=0.4$~GeV), respectively (see Fig.~\ref{fig:fragbs}).  
Convoluting  this FFs with $b\bar b$ production cross section in LO of QCD we obtain the $p_T$ dependence of ${f_{B_s}}/{f_{B}}$ showed in  Fig.~\ref{fig:BsBd}, which is close to one, that was experimentally obtained by the LHCb experiment~\cite{Aaij:2013qqa} and shown in  Fig.~\ref{fig:BsBd_LHCb} (see also~\cite{Aaij:2011hi}).  We need to stress here, that the chosen parameters of the FFs for  $B$ and $B_s$ mesons were not obtained from the experimental data.  However, it could be concluded, that the LHCb data on 
${f_{B_s}}/{f_{B}}$ could be described within the fragmentation model. 
Of course we should keep in mind, that the application of the  fragmentation approach is not well-motivated at low transverse momenta. 

Comparing Fig.~\ref{fig:BsBd} and  Fig.~\ref{fig:BsBd_LHCb} one can conclude that the linear approximation used by the LHCb Collaboration is not the best choice to describe  the  $f_s/f_d$ dependence as function of $p_T$.

\section{Recombination}

\begin{figure}[!t]
\centering
\resizebox*{1.0\textwidth}{!}{
\includegraphics{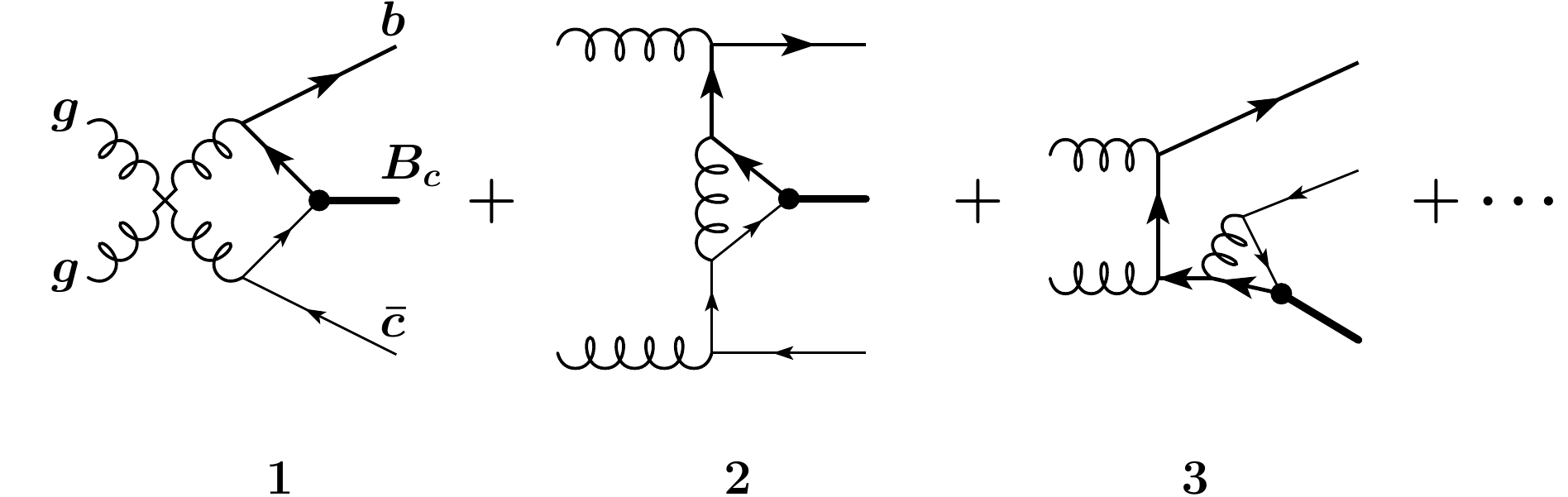}
\hfill
\includegraphics{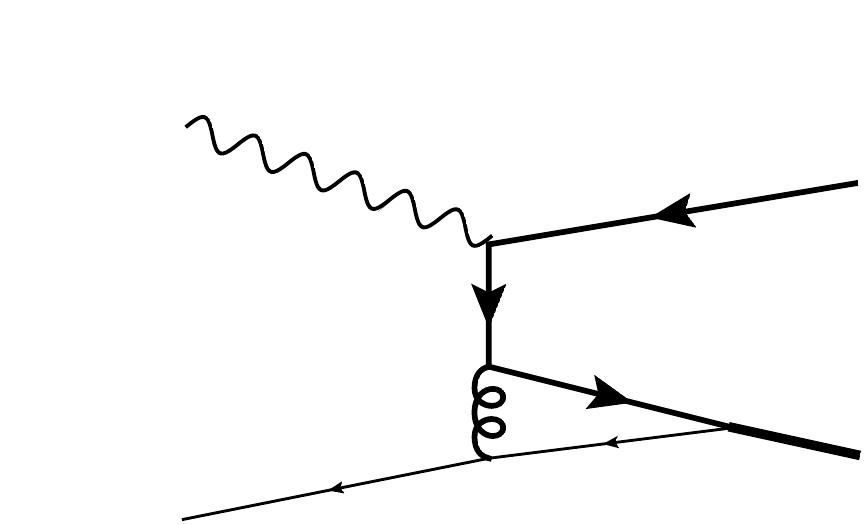}
}
\\
\parbox[t]{0.47\textwidth}{
\caption{The examples of diagrams for the process $gg\to B_c +X$.}
\label{fig:gg_Bc_diagr}   \hfill}
\hfill
\parbox[t]{0.47\textwidth}{
\caption{The production of heavy-light quark pair in  the quark-photon interaction.} 
\label{fig:qphoton_diagr}  \hfill}
\end{figure}

\begin{figure}[!t]
\centering
\resizebox*{1.0\textwidth}{!}{
\hspace*{-0.5cm}\includegraphics{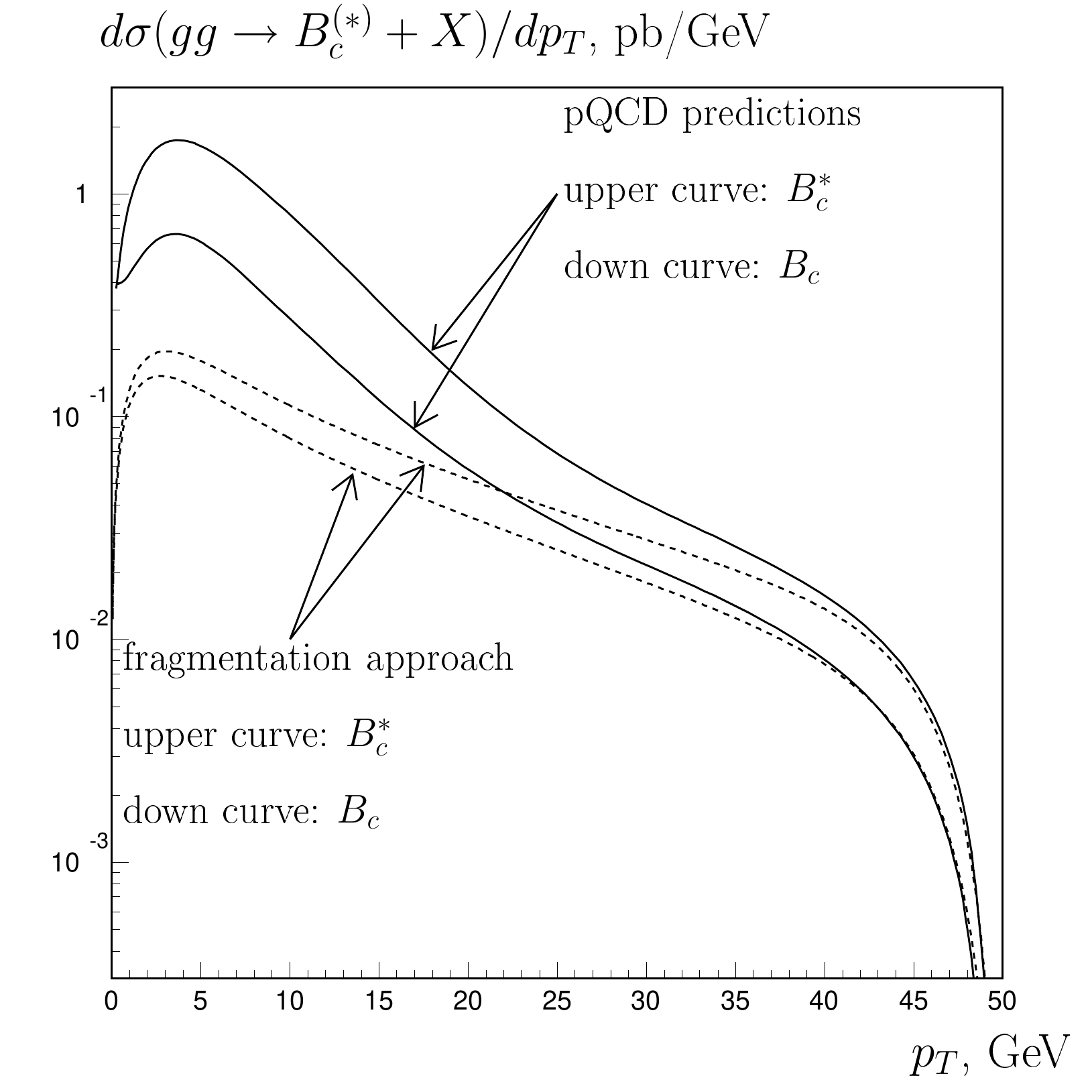}
\hfill
\includegraphics{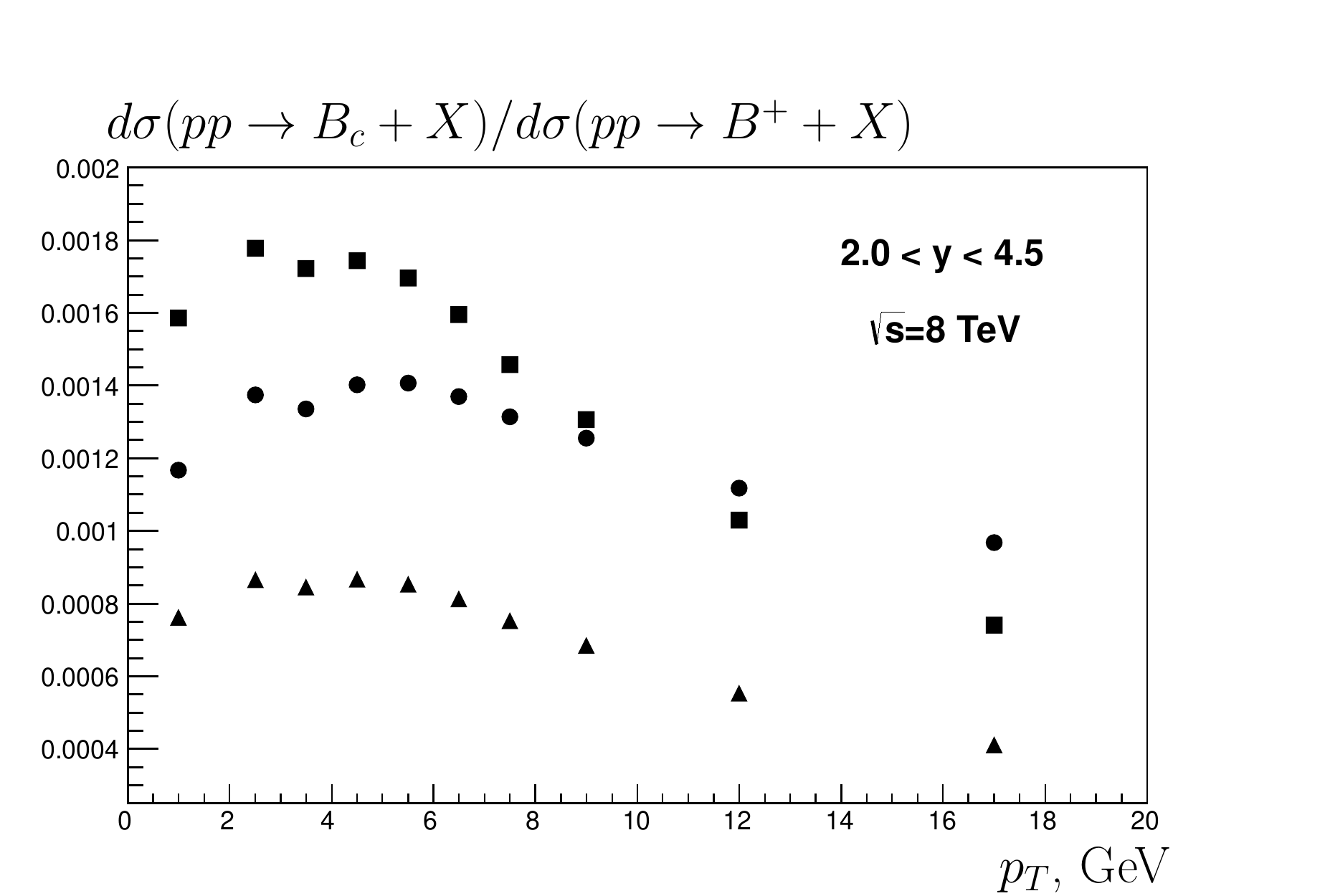}
}
\\
\parbox[c]{0.47\textwidth}{
\caption{The $p_T$ distribution of gluonic $B_c$ production cross section at interaction energy 100 GeV.}

\label{fig:gg_Bc_Pt}}
\hfill
\parbox[c]{0.47\textwidth}{
\caption{$f_{B_c}/f_{B^+}$ ratio as function of $p_T$. Several scale choices have been used for the $B_c$ yield calculation: $\mu_R=\mu_F=10$~GeV (black circles);
$\mu_R=\mu_F=\sqrt{s_{gg}}/4$ (black squares);  $\mu_R=\sqrt{s_{gg}}/4$,  $\mu_F=\sqrt{s_{gg}}/2$ (black triangles). }
\label{fig:BcBd}  \hfill}
\end{figure}

The most clear example, where the main contribution is provided by the non-fragmentation  mechanism is the hadronic production of $B_c$ mesons.  It was shown in~\cite{Berezhnoy:1994ba,Berezhnoy:1995au,Berezhnoy:1997fp}, that due to the diagrams of non-fragmentation (recombination) type the production can not be represented as a $b$-quark production followed by fragmentation up to $p_T\sim 35\div 40$~GeV (see Fig.~\ref{fig:gg_Bc_diagr} and \ref{fig:gg_Bc_Pt}).

It can be shown that the diagrams of types 1  and 2 (amplitudes $A_1$ and $A_2$, correspondingly) mostly contribute at small and medium transverse momenta and their contributions decrease with increasing $p_T$ as $1/ p_T^6$:
$$|A_1|^2, \; |A_2|^2 \sim \frac{1}{p_T^6},$$
 
 whereas contributions of the fragmentation type diagrams (amplitudes $A_3$) behave like $1/ p_T^4$:
 
  $$|A_3|^2 \sim \frac{1}{p_T^4}.$$

It is worth to note that the set of fragmentation diagrams and the set of recombination diagrams are not separately gauge invariant, and therefore interfere.  In this sense, the separation on the fragmentation and recombination mechanism is quite conditional.

In Fig.~\ref{fig:BcBd} we predict the shape for $f_{B_c}/f_B$ ratio.
The $B_c$ meson  yield has been estimated  using the cross section for the subprocess  $gg\to B_c + X $ calculated within LO pQCD for several scale choices. The interaction with a sea $c$-quark  ($g c_\mathrm{sea}\to B_c + X $) has been neglected.
 The $B$-meson cross section has been obtained within FONLL~\cite{Cacciari:1998it,Cacciari:2012ny}\footnote{ The calculation within FONLL has been done by us with the help of interactive web-form http://www.lpthe.jussieu.fr/$\sim$cacciari/fonll/fonllform.html. Therefore the  FONLL group members are not responsible for the obtained values. To estimate $f_{B_c}/f_{B^+}$ ratio we have used "central" values of the $B$-meson cross section.}.  Quantitatively we can not guarantee $p_T$ dependence of $f_{B_c}/f_{B^+}$.  But we are confident that these predictions for the ratio shape are qualitatively correct, and there are reasons to expect a quite rapid decrease of this ratio with increasing $p_T$, and do not expect to see the plateau at least at $p_T < 25$~GeV.

It was shown in~\cite{Braaten:2001uu,Berezhnoy:2005dt,Berezhnoy:2000ji,Berezhnoy:1999yj}, that the recombination contribution also can play an essential role in the photoproduction of heavy-light mesons. Let us suppose that we use a heavy-light pair $Q\bar q$ in color singlet-state   instead of a single heavy quark as a pattern to construct a heavy-light meson. Such  pairs can be produced in the quark-photon interaction $\bar q \gamma \rightarrow (Q \bar q) +\bar Q$, where the light valence quark of the formed meson is produced already in the hard subprocess. Thus it is not fragmentation, it is a recombination. As it was demostrated in~\cite{Braaten:2001uu}  this recombination contribution is not suppressed in the forward region and decreases with increasing $p_T$ as $1/ p_T^6$:
\begin{eqnarray}
\left.
\frac{d \hat{\sigma}[\bar q \gamma \rightarrow (Q \bar q) +
\bar Q]}{ d \hat{\sigma}[g \gamma \rightarrow Q \bar Q]}\right|_{\theta = \pi/2}
\approx  \frac{256 \pi}{ 189} \alpha_s \frac{m_Q^2}{m_Q^2+p_T^2},
\end{eqnarray}

\begin{equation}
\frac{d \hat{\sigma}[\bar q \gamma \rightarrow (Q \bar q) +
\bar Q]}{dp_T} \sim \frac{1}{p_T^6},
\end{equation}
\begin{eqnarray}
\left.
\frac{d \hat{\sigma}[\overline{q} \gamma \rightarrow (Q \bar q) +
\bar Q]}{ d \hat{\sigma}[g \gamma \rightarrow Q \bar Q]}\right|_{\theta = 0}
\approx  \frac{256 \pi}{ 81} \alpha_s \sim 1.
\end{eqnarray}

Therefore, the recombination could contribute essentially not only into doubly heavy meson production, but also into heavy-light meson production.

\section{Conclusions}

 In  this research we show, that the the dependence of $f_s/f_d$ on $p_T$ observed by LHCb~\cite{Aaij:2013qqa} could be understood within the fragmentation approach. However one should keep in mind, that
the validity of the fragmentation model is questionable at low $p_T$.   Contributions of nonfragmentation mechanisms are also possible. The precise measurements of $f_s/f_d$, $f_{\Lambda_b}/f_{B}$ and analogous values are needed at all $p_T$  to prove or reject the existence of nonfragmentation contribution into heavy hadron production.

 A plateau is expected  in  $f_s/f_d$  distribution at high $p_T$.

The value of $f_{B_c}/f_B$ should  decrease  with increasing $p_T$. A plateau at high $p_T$ could exist for 
$f_{B_c}/f_B$, but, seems, to be unreachable for the LHC experiments.

We would like to thank Vanya Belyaev and Niels Tuning for the fruitful discussion.

\bibliography{/home/aber/papers/heavy_quark}
\end{document}